# An Extensive Comparison of Static Application Security Testing Tools

Matteo Esposito
m.esposito@ing.uniroma2.it
University of Rome "Tor Vergata"
Rome, Lazio, Italy

Valentina Falaschi
valentina.falaschi@alumni.uniroma2.eu
University of Rome "Tor Vergata"
Rome, Lazio, Italy

Davide Falessi
falessi@ing.uniroma2.it
University of Rome "Tor Vergata"
Rome, Lazio, Italy

## ABSTRACT

**Context:** Static Application Security Testing Tools (SASTTs) identify software vulnerabilities to support the security and reliability of software applications. Interestingly, several studies have suggested that alternative solutions may be more effective than SASTTs due to their tendency to generate false alarms, commonly referred to as low Precision. **Aim:** We aim to comprehensively evaluate SASTTs, setting a reliable benchmark for assessing and finding gaps in vulnerability identification mechanisms based on SASTTs or alternatives. **Method:** Our SASTTs evaluation is based on a controlled, though synthetic, Java codebase. It involves an assessment of 1.5 million test executions, and it features innovative methodological features such as effort-aware accuracy metrics and method-level analysis. **Results:** Our findings reveal that SASTTs detect a tiny range of vulnerabilities. In contrast to prevailing wisdom, SASTTs exhibit high Precision while falling short in Recall. **Conclusions:** Our findings suggest that enhancing Recall, alongside expanding the spectrum of detected vulnerability types, should be the primary focus for improving SASTTs or alternative approaches, such as machine learning-based vulnerability identification solutions.

## 1 INTRODUCTION

Our daily life relies on software that carries out everyday tasks [78], from our household [5] to autonomous driving [30] and digital governance [12]. Vulnerabilities are causing significant financial losses to businesses [61] and threatening critical security infrastructures [62]. The outspreading of cyber-warfare exacerbates the importance of preventing vulnerabilities [1].

Static Application Security Testing Tools (SASTTs) are a white box test method that analyses application code, without running it, for detecting security vulnerabilities in application source code [15]. SASTTs use syntax, data flow, control flow, or their combination [15].

The Common Weakness Enumeration (**CWE**), created by MITRE [40], provides a comprehensive list of vulnerabilities concerning software systems. CWEs are independent of the programming language and provide detailed information about the vulnerability type, such as a unique identifier, description, related attack patterns, and references to additional information. The Juliet test case suite JTS [29], developed by the National Institute of Standards and Technology [44], is a repository of code fragments written in several programming languages, specifically developed to include or not include a specific CWE. JTS is a reliable benchmark for testing tools that identify vulnerable code, including, but not limited to, SASTTs.

Many previous studies [63, 79, 16, 54, 74, 66, 76, 77] motivated the use of solutions that are alternatives to SASTTs, e.g., machine learning based techniques, by claiming that SASTTs produce many false alarms. In conclusion, it is important to set a comprehensive benchmark for assessing and enhancing the effectiveness on vulnerability identification mechanisms based on SASTTs or based on alternatives such as machine learning.

This paper aims to rigorously and extensively evaluate SASTTs accuracy. Specifically, we investigate the following research questions:

**RQ 1.** *What is the extent of SASTTs coverage?*: Given the high number of SASTTs and CWEs, it is crucial to understand which SASTT can identify which CWEs. Specifically, in this research question, we investigate which SASTT claims to identify which CWEs, if there is a difference between the claims and the identified CWEs, the overlap across SASTTs in identifying CWEs, and finally, which CWEs is identified by no SASTT.

**RQ 2.** *How accurate are SASTTs?*: Since multiple SASTTs identify the same CWE, evaluating and comparing SASTTs accuracy supports SASTTs selection and improvements. Specifically, in this research question, we investigate the differences in SASTTs accuracy between the CWEs expected to be identified by a SASTT versus the CWE actually identified by the SASTT, and if the accuracy varies across SASTTs. As SASTTs identify different sets of CWEs, a comparison of SASTTs accuracy might be biased by characteristics of the identified CWEs such as the intrinsic difficulty of, and the proportion of positive test cases of, test cases in JTS. To mitigate this bias, for each CWE and accuracy metric, we select the most accurate SASTT; afterwards, we analyse if the best SASTT is statistically better than other SASTTs and if it varies across accuracy metrics.

In this work we use the term *accuracy* to refer specifically to the performance of SASTTs overall, i.e., in a broader meaning than specific IR accuracy aspects such as precision and recall.
Our evaluation of SASTTs features the following novelties over previous SASTTs evaluations [28, 48, 34, 13, 26, 4, 2, 43, 50, 32, 80, 52, 27, 82, 63, 79, 16, 54, 39, 9] on JTS:

- largest set of considered SASTTs,
- largest set of considered CWEs,
- analysis of uncovered CWEs and test cases,
- actual vs expected analysis,
- method-level analysis,
- effort-aware analysis [10].

Our results related to eight SASTTs, 112 CWEs, and 183.769 (183k) total test cases, totalling 1,411,608 test executions, provide many relevant insights to practitioners and researchers for improving software security testing. One of the most important results is that our SASTTs have been unable to identify most vulnerability types despite each type being represented by hundreds of test cases. Moreover, our study shows that SASTTs excel in Precision while falling short in Recall; i.e., false negatives are more than false positives.

The remainder of the paper is structured as follows. We present the research background and related work in Section 2. We describe the study design in Section 3. Section 4 discusses the empirical results. We discuss the results in Section 5. We explain the threats to the validity of our study in Section 6. Finally, we provide our conclusions and future works in Section 7.

## 2 BACKGROUND & RELATED WORK

In this section, we present the background and the related work using SASTTs. Undiscovered vulnerabilities can lead to costly impacts on software firms [61]. Cavusoglu et al. [11] show that a fast fix is essential to avoid loss connected to the vulnerability and its public disclosure. Louridas [36] presents the considerable cost associated with inadequate software testing infrastructure; thus, a vulnerability discovery tool is needed [17]. SASTTs are a practical choice to actively discover software vulnerabilities during development [8, 55, 7, 81]. The National Vulnerabilities Database NVD [45] collects the Common Vulnerabilities and Exposures (**CVE**) catalogued via the CWEs [38]. NVD also classify the vulnerability severity associated with the specific CVEs [20]. SASTTs are developed to target CWEs for vulnerability discovery. While multiple SASTTs can identify the same vulnerability, only a few can detect more specific and challenging CWEs [48]. NIST provides the SARD Test Suites [6] targeting the most common Programming Languages for benchmarking SASTTs. SASTT analysis can support code review. Tufano et al. [65] most recently conducted an exhaustive examination of the state of the art in code review. The authors evaluate ChatGPT's effectiveness and found that it falls short of outperforming human expertise in code review. Therefore, human intervention remains necessary, regardless of the extent of automation implemented.

### 2.1 Static application security testing tools

The wide spreading of software vulnerability increases over time, and new vulnerability discovery is becoming a matter of every day [42]. To aid developers in reducing the vulnerability present in the latest releases of software, various SASTTs exist [4, 48, 43, 68] Those SASTTs target different and specific vulnerabilities, hence showing diverse capabilities among [13]. The most important and used SASTTs [48, 34, 13, 26, 4, 2, 43, 32] include the following:

FindSecBugs: Security plugin for SpotBugs analyzing Java code for vulnerabilities [25]. Infer: Facebook's scalable static analysis tool for Java, C, and C++ detecting errors and vulnerabilities [21]. JLint: Open-source Java code analyzer for identifying errors and security issues [3]. PMD: Java static analysis tool providing detailed reports on issues like unused code [49]. SonarQube: Open-source code quality management tool supporting multiple languages [59]. Snyk Code: Identifies code issues in Java and Python projects, offering detailed reports [58]. Spotbugs: Free Java bytecode analyzer for problems and vulnerabilities with bytecode analysis [60]. VCG: Security review tool for various languages, including Java, with customizable checks [67].

Yang, Tan, et al. [74], discusses the challenges associated with SASTTs in software development due to difficulties handling large numbers of reported vulnerabilities, a high rate of false warnings and a lack of guidance in fixing reported vulnerabilities. To address these challenges, the authors propose a set of approaches called Priv. The authors shows that it effectively identifies and prioritises vulnerability warnings and reduces the rate of false positives.

### 2.2 Static application security testing tools evaluation on test suites

Table 1 presents the critical aspect of related work on evaluating SASTTs on publicly available benchmarks such as JTS. We selected the most recent and relevant work at the time of writing. Oyetoyan et al. [48] discuss the inclusion of tools to aid agile development pipeline in focusing on security aspects; the authors choose to target the 112 CWE provided by the JTS (v. 1.2) testing a total of six different tools, focusing on the accuracy metrics. Moreover, the authors provided the scripts for replicate the study and the data. Li et al. [34] presents the challenge of using IDE-integrated plugins for vulnerability discovery; the authors target 29 CWE from the Open Worldwide Application Security Project (OWASP) 2017 TOP 10 [47], focusing on accuracy and usability [57] metrics, testing a total of five different tools. Li et al. [34] highlighted the differences in what the SASTTs claim to be able to discover and what they can; in the same vain, our work discusses this issue in Section 4. Charest et al. [13] discuss the importance of the SASTTs and their actual discovery capabilities; the authors compared four open-source tools on four specific CWEs from the JTC, focusing on java methods rather than classes. Goseva-Popstojanova et al. [26] compared SASTTs targeting Java and C++ programming languages resulting, in some cases, in performance worse than the random flip of a coin; the authors tested three SASTTs on 41 CWEs belonging to JTC and Open-Source Software (OSS), targeting functions an methods and focusing on accuracy metrics. Arusoaie et al. [4] investigate on the accuracy of custom developed SASTTs for detecting C++ vulnerability; the authors tested 12 SASTTs targeting 51 CWE from the

Table 1: Related work on the usage of SATTs on JTS.

| | Oyetoyan et al. [48] | Li et al. [34] | Charest et al. [13] | Goseva-Popstojanova et al. [26] | Arusoaie et al. [4] | Alqaradaghi et al. [2] | Nguyen-Duc et al. [43] | Our Work |
|---|---|---|---|---|---|---|---|---|
| **Granularity** | File | File * | Method | Functions / Methods | File | Method | File | Method |
| **PL** | Java | Java | Java | Java, C++ | C++ | Java | Java | Java |
| **CWE** | 112 | OWASP [47] | 4 | 22+19 | 51 | MITRE [41] | 112 | 112 |
| **Aspects** | Accuracy, Interview | Accuracy, Usability | Accuracy | Accuracy | Accuracy | Accuracy | Accuracy, Interview | Accuracy |
| **Dataset** | Juliet v1.2 | Juliet | Juliet v1.2 | Juliet, OSS | Toyota [64] | Juliet v1.3 | Juliet v1.3 | Juliet v1.3 |
| **#Tool** | 5+1 | 5 | 4 | 3 | 12 | 4 | 7 | 8 |
| **Results** | No | No | No | No | No | No | **Yes** | **Yes** |
| **Scripts** | **Yes** | No | No | No | **Yes** | No | No | **Yes** |
| **Datasets** | **Yes** | No | No | No | **Yes** | No | **Yes** | **Yes** |

Toyota ITC Test Suite [56] focusing on accuracy metrics such as detection rate and false positive rate. Moreover, the authors provided the scripts for replicate the study and the data. Alqaradaghi et al. [2] conduct a comparison experiment with open source SASTTs targeting specific CWE highliting differences in SASTTs ability to discover certain CWEs; the authors targeted six of the official "Weaknesses in the 2020 CWE Top 25 Most Dangerous Software Weaknesses" [41], testing four open source SASTTs focusing on accuracy metrics. Nguyen-Duc et al. [43] investigate the combination of SASTTs results to improve the final result; the authors targeted the 112 CWE from the JTT, testing seven SASTTs, focusing on files and delivering a positive result if and only if all SASTTs agreed on positive. Moreover, the authors provided the raw results and the data. Li et al. [34] perform a comprehensive study on using five IDE plugins to report vulnerability issues. The authors identify a disparity between the advertised detection capabilities and the actual detection performance of the five plugins, resulting in a high false positive rate.

Finally, our work focuses on discovering vulnerable methods [26, 2] targeting 112 CWEs from the JTT for Java, testing six open source SASTTs and two commercial ones. We provide the raw results, the scripts for replicate the study and the data.

## 3 METHODOLOGY

The following section outlines our design decisions and provides the rationale behind them to address our research questions. We detail the tools employed, our measurement process, and the statistical tests conducted.

### 3.1 RQ 1. *What is the extent of SASTTs coverage?*

Given the substantial number of available SASTTs and CWEs, it becomes crucial to determine which SASTT accurately detects particular CWEs and which ones go undetected.

We define coverage as the percentage of CWEs identified, at least once, by one or more SASTTs. To understand the expected coverage, we assess the SASTT's documentation claims regarding its ability to identify specific CWEs. To gauge the actual coverage, we evaluate the SASTT's performance in identifying specific CWEs. A SASTT may claim or actually identify a CWE. Therefore, a combination of SASTT and CWE can fall into one of the following statuses:

- *Not Expected*: The SASTT does not claim, in the documentation, to be able to identify the CWE.
- *Expected (ECWE)*: The SASTT claims, in the documentation, to identify the CWE. If expected, the combination of SASTT and CWE can be in two sub-statuses:
  - *Actual (ACWE)*: The SASTT identifies the CWE in at least one test case; i.e., there is at least one true positive.
  - *Not Actual*: The SASTT identifies the CWE in no test case; i.e., there is no true positive.

To analyze the distribution of ECWEs and ACWEs, we employ UpSet graphs [14]. UpSet is a graphical visualization technique describing the intersection of datasets [33]. The dimension of a rectangle represents the cardinality of the set. In our context, a set is the specific group of CWEs expected or actually identified by a specific group of SASTTs.

We also investigate how the number of unique ECWEs or ACWEs increases with the number of SASTTs. This helps us understand how many SASTTs are worth using if we want to cover a specific number of CWEs.

### 3.2 RQ 2. *How accurate are SASTTs?*

Since multiple SASTTs can identify the same CWEs, it is important to understand the accuracy of SASTTs.

Our dependent variable is the accuracy of SASTTs measured by four accuracy metrics. Three of our accuracy metrics are standard in information retrieval:

- **Precision**: $\frac{TP}{TP+FP}$.
- **Recall**: $\frac{TP}{TP+FN}$.
- **F1-score**: $\frac{2*Precision*Recall}{Precision+Recall}$.

Effort-aware metrics (EAMs) relate to the ranking, in this context provided by a SASTTs, of candidate vulnerable entities [10]. The better the SASTTs, the higher the number of vulnerable entities a developer can identify within a given rank. The rationale behind EAMs is that they focus on effort reduction gained by using SASTTs [10]. We adopt one EAM: NPofB20 [10]. NPofBX is the percentage of defective entities discovered by examining the first x% of the code base, ranked by their probabilities, to be vulnerable, divided by their size, i.e., LOC. For example, a method having an NPofB20 of 30% means that, following the ranking, we can detect 30% of vulnerable entities by examining only 20% of the codebase. Thus, NPofB20 is the percentage of vulnerable entities discovered by examining the first 20% of the code base, ranked by their probabilities, to be vulnerable, divided by their size. The value for $x = 20\%$ is commonly used in the state-of-the-art as a cutoff value to set the effort required for the defect inspection [10, 51, 75, 46, 72, 73, 23, 19].

In this paper, we posit the following null hypotheses:

(1) H01: There is no difference in the accuracy of SASTTs between ECWEs and ACWEs.
(2) H02: There is no difference in the accuracy of SASTTs.
(3) H03: The best SASTT across CWEs has the same accuracy as other SASTTs.

For each SASTT, we first compare the accuracy between ECWEs and ACWEs. This is done to understand how reliable past comparisons of SASTTs focusing on ECWEs rather than ACWEs are. Afterwards, to compare the accuracy of SASTTs, we consider ACWEs only.

Since different SASTTs observe different CWEs, a standard comparison of SASTTs based on the accuracy of their ACWEs can be biased by the characteristics of their ACWEs, such as the number of test cases and intrinsic difficulty level. To mitigate the impact of the cardinality of test cases on our SASTTs comparison, we analyze the accuracy per CWE and the weighted average accuracy across CWEs.

We test H01 and H02 via the Wilcoxon signed-rank test [70]. The Wilcoxon signed-rank test (**WT**) is a non-parametric statistical test that compares two related samples or paired data. WT uses the absolute difference between the two observations to classify and then compares the sum of the positive and negative differences. The test statistic is the lowest of both. We select WT to test H01 and H02 because the four SASTTs' accuracy metrics resulted as not normally distributed; hence, we use the Wilcoxon test instead of the paired t-test, which assumes a normal data distribution.

We set our alpha to 0.01. We decided to reduce alpha from the standard value of 0.05 due to the many statistical tests we perform and hence to reach a good balance between type 1 and type 2 error [31].

To test H03, we use Dunn's all-pairs test [18]. The Dunn All-Pairs Test, also known as the Dunn Test (**DT**) or Dunn-Bonferroni Test, is a post hoc test used for statistical analysis. It is a non-parametric test that compares the differences between all group pairs in a dataset. DT compare the rank sums of all possible group pairs and obtains a p-value set for each paired comparison. These p values are adjusted with Bonferroni corrections to mitigate type I errors. The adjusted p-values are compared to a pre-defined significance level to determine the statistically significant comparison pairs. We select DT to test H03 because the four SASTTs' accuracy metrics are not normally distributed, and unlike WT, DT is a pair-wise test.

### 3.3 The Dataset

We analyze eight SASTTs: SpotBugs 4.6, FindSecBugs 1.12, VCG 2.2, PMD 6.44, SonarQube 9.5, Infer 1.1, Jlint 3.1.2, and Snyk 1.1073. For further details about the SASTTs, please see Section 2.1.

To evaluate SASTTs, we use the Java version of the JTS V. 1.3, its purpose is to evaluate the capabilities of SASTTs specifically. The version used in the present paper includes 112 CWEs. Each CWE has a different number of test cases; each test case is a method in Java that is defined to contain or not contain a specific CWE. Specifically, each CWE has a set of Java classes; each Java class has multiple methods implementing the same logic. Each test case is named "goodXY" if it implements logic Y without CWE X or "badXY" if it implements logic Y with CWE X. The total number of test cases in JTS is 176,451. To illustrate the concepts above, Listing 1 presents a simple example of a "good" method for CWE 209 Listing, and 2 presents a simple example of the "bad" method. CWE-209, i.e., Generation of Error Message Containing Sensitive Information, refers to a security weakness where an application reveals sensitive information through error messages displayed to users or logged in a file. When an application displays an error message, it should provide only the necessary information to help users understand the cause of the error and how to resolve it. However, suppose an error message contains sensitive information, such as passwords, private keys, stack traces or other confidential data. In that case, attackers can exploit it to grasp the info of source files.

Figure 1 reports the distribution of the number of test cases across CWEs in JTS. According to Figure 1, the number of test cases for CWE ranges within 1 (CWE 111) and 29,095 (CWE 190), with an average value of 1.250 across CWEs.

### 3.4 Measurement Procedure

Figure 2 presents our measurement procedure. We set up the environment requested by each SASTT. We then performed the measurement process in five steps: Since SASTTs report violations without explicitly referring to specific CWEs, in *Step 1* we manually map each SASTT reported violation to its specific CWE developing the Tool Mapping. In *Step 2* we scan the JTS with each SASTT which produces the Juliet Test Suite Results. Afterwards, in *Step 3* we develop the Parsed Output by the Juliet Test Suite Results with the Tool Mapping. Thereafter, in *Step 4* we measure the accuracy of the Parsed Output by comparing the results against the JTS documentation [53] obtaining the *Juliet accuracy report*. Subsequently, in *Step 5* we interpret the SASTTs results according to Algorithm 1 to get the *confusion matrix*. Specifically, we computed the confusion matrix elements as follows:

- *True Positive*: the test case is defined as affected by CWE and the SASTT identifies it.
- *True Negative*: the test case is defined as not affected by CWE and the SASTT doesn't identify it.
- *False Positive*: the test case is defined as not affected by CWE and the SASTT identifies it.
- *False Negative*: the test case is defined as having a CWE and the SASTT doesn't identify it.

Afterwards, we identify the ACWEs for a SASTT (see Section 3.1) as the ECWEs having at least a true positive for the SASTT.

## 4 RESULTS

This section aims to answer the research questions and test the hypotheses posed in the design section.

```java
public void bad() throws Throwable {
    try
    {
        throw new UnsupportedOperationException();
    }
    catch (UnsupportedOperationException exception)
    {
        // FLAW: Print stack trace to console on error
        exception.printStackTrace();
    }
}
```

Listing 1: Example of a method affected by CWE 209.

```java
private void good1() throws Throwable{
    try
    {
        throw new UnsupportedOperationException();
    }
    catch (UnsupportedOperationException exception)
    {
        // FIX: print a generic message
        IO.writeLine("Unsupported operation error");
    }
}
```

Listing 2: Example of a method not affected by CWE 209.

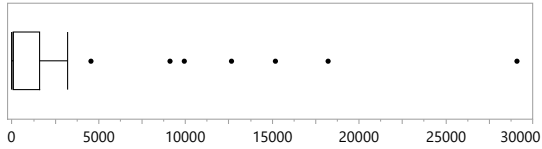

Figure 1: Distribution of number of test cases across CWEs.

Table 2: Number of ECWEs and ACWEs per SASTT.

|  | T1 | T2 | T3 | T4 | T5 | T6 | T7 | T8 |
|---|---|---|---|---|---|---|---|---|
| **Expected** | 58 | 19 | 7 | 6 | 26 | 18 | 8 | 4 |
| **Actual** | 9 | 11 | 7 | 6 | 12 | 12 | 6 | 3 |

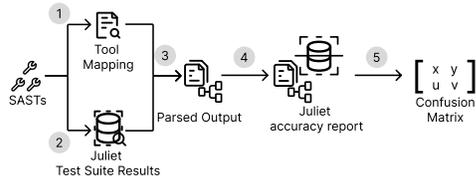

Figure 2: Workflow of the measurement procedure.

---

**Algorithm 1** Confusion matrix value.

**if** FileName$MethodName contains 'bad' **then**
　**if** cwe_predicted == null || cwe_predicted ≠ cwe_id **then**
　　predicted ← no
　　output False Negative
　**if** cwe_predicted == cwe_id **then**
　　predicted ← yes
　　output True Positive
**else**
　**if** cwe_predicted == null || cwe_predicted ≠ cwe_id **then**
　　predicted ← no
　　output True Negative
　**if** cwe_predicted == cwe_id) **then**
　　predicted ← yes
　　output False Positive
**if** cwe_predicted == (?) **then**
　predicted ← ?
　output ?

---

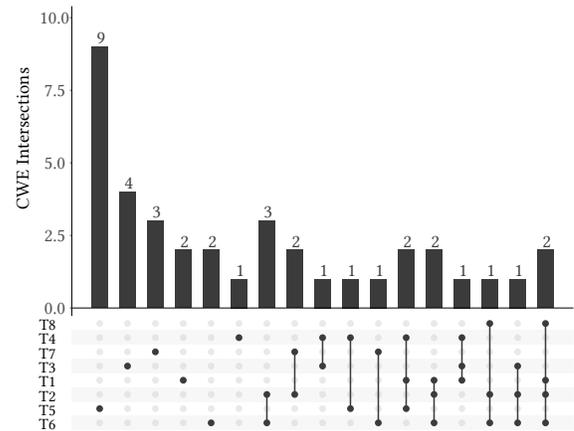

Figure 3: UpSet graph on the ACWEs per SASTT.

## 4.1　RQ 1.　*What is the extent of SASTTs coverage?*

We investigate the SASTTs coverage in terms of ECWEs and ACWEs. Table 2 presents the distribution of ECWEs over our eight SASTTs. According to Table 2, the number of ECWEs and ACWES varies between four (T8) and 58 (T1). Moreover, we note that the total number of non-unique ECWEs across our eight SASTTs is 146. Table 2 presents the number of ACWEs and ECWEs per SASTT. According to Table 2, the ACWEs are less than half of the ECWEs in total and for most of SASTTs. Moreover, two SASTTs identify all ECWEs, i.e., ACWEs == ECWEs.

Figure 3 presents the UpSet intersection graph of ACWEs across SASTTs. There is no CWE that is identified by more than four SASTTs. Moreover, six out of eight SASTTs have at least one exclusive ACWE. According to Figure 3, T1 has only two exclusive ACWEs, i.e., only 10% of its exclusive ECWEs. Furthermore, T5 has the highest number of exclusive ACWEs, i.e., 9. Finally, it is interesting that most ACWEs relate to only a single SASTTs.

Table 3 presents the increment in the number of ECWEs and ACWEs as the number of SASTTs increase. Table 3 shows that the actual and expected coverage does not increase over six SASTTs. Specifically, our set of eight SASTTs expects to identify 70% of CWEs in JTS but, according to our results, they actually identify only 34% (see Table 3, column 6). Thus, most of the CWEs are expected but not actually identified and two-thirds of CWEs are not identified by any SASTT. Finally, we note that a single SASTT covers a maximum of 12 ACWEs, i.e., 11% of JTS, and 58 ECWEs (i.e., 52% of JTS).

Table 3: Max unique ACWEs and CCWEs by SASTT and JTS Coverage (%).

| No. of SASTTs | 1 | 2 | 3 | 4 | 5 | 6 | 7 | 8 |
|---|---|---|---|---|---|---|---|---|
| ECWEs | 58 (52%) | 68 (60%) | 72 (64%) | 76 (67%) | 78 (69%) | 79 (70%) | 79 (70%) | 79 (70%) |
| ACWEs | 12 (11%) | 24 (21%) | 30 (27%) | 35 (31%) | 37 (33%) | 38 (34%) | 38 (34%) | 38 (34%) |

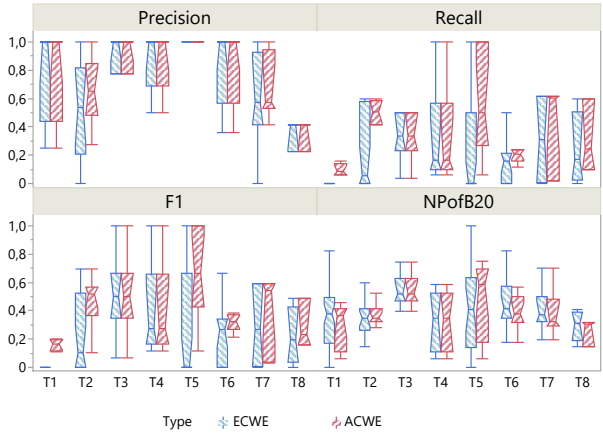

Figure 4: Distribution of SASTTs accuracy between ECWEs and ACWEs.

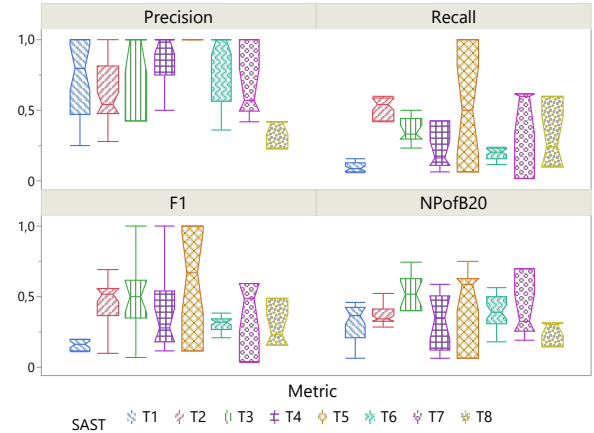

Figure 5: Distribution of SASTT accuracy on ACWEs.

Table 4: Accuracy gain of SASTTs on ACWEs over ECWEs, i.e., H01 test results. ∗ = p-value < 0.001

|  | T1 | T2 | T3 | T4 | T5 | T6 | T7 | T8 |
|---|---|---|---|---|---|---|---|---|
| **Precision** | 0% | 0% | 0% | 0% | 0% | 0% | 162% ∗ | 0% |
| **Recall** | 156% ∗ | 12% ∗ | 0% | 0% | 237% ∗ | 4% ∗ | 501% ∗ | 0% |
| **F1** | 156% ∗ | 12% ∗ | 0% | 0% | 237% ∗ | 4% ∗ | 501% ∗ | 0% |
| **NPofB20** | -14% ∗ | -3% ∗ | 0% | 0% | 20% ∗ | -1% | -24% ∗ | 0% |

## 4.2 RQ 2. *How accurate are SASTTs?*

Regarding the accuracy of ECWEs versus ACWEs, Figure 4 provides the distributions across ECWEs versus ACWEs, of each SASTTs accuracy metric. According to Figure 4, the difference between ECWEs and ACWEs is more evident in Recall than Precision. Specifically, the median Recall of ACWEs is higher than ECWEs in all SASTTs. Interestingly, T1 shows zero Recall for ECWEs and a positive Recall for ACWEs. In six out of eight SASTTs, ACWEs exhibit higher F1 scores than ECWEs. There is no observable difference in NPofB20 between ACWEs and ECWEs.

Table 4 reports the statistical test results about the differences, and the gain, in our four accuracy metrics, between ECWEs and ACWEs of SASTTs. We report the gain as statistically significant (i.e., p-value < 0.0001) with an asterisk (*). According to Table 4, we can reject H01 in 15 out of 32 cases. Moreover, we can reject H01 for all T7 four accuracy metrics. Considering only the ACWEs rather than all ECWEs significantly increases the Recall of most SASTTs; in particular, the Recall increases by more than 150% for three SASTTs (T1, T5, and T7). Interestingly, in four out of eight SASTTS, NPofB20 exhibits a negative and statistically significant variation. Consequently, ACWEs are ordered in a manner unfavourable to NPofB20.

Regarding the accuracy of SASTTs, Figure 5 reports the accuracy metrics per SASTTs on all the ACWEs. According to Figure 5, in all eight SASTTs, the median Precision across ACWEs, is higher than the median Recall. Moreover, T3, T4, T5, and T6, show a perfect median Precision; thus, most of the SASTTs provide no false positive. We note that five out of eight SASTTs show a median Recall below 0.5; moreover, 11 out of our 38 ACWEs, i.e., 29%, have a maximum Recall lower than 0.5. In about a third of ACWEs, all SASTTs provide more false negatives than true positives, i.e., the number of undetected vulnerabilities is higher than the detected ones. Finally, regarding NPofB20, by ordering the test cases and inspecting only the first 20% we would discover, on average, at least 30% of positive cases.

Table 5 presents the weighted average accuracy, across test cases. T5 shows the maximum average across tests and the highest Recall across ACWEs. However, according to Figure 5 the Recall, F1 and NPofB20 distributions of T5 are very wide, ranging from 0.08 to 1.0. Thus, even the most accurate SASTT in Recall misses many test cases.

Regarding comparing the accuracy across SASTTs, p-values resulted lower than $\alpha$ in all four accuracy metrics and hence we can reject H02 and claim that SASTTs differ in accuracy.

Regarding how the best SASTT compare to other SASTTs, according to Table 5, our best SASTT is T5. According to the DT results comparing differences in the accuracy between T5 and other SASTTs, we can reject H03 in 26 out of 28 cases. Thus, in all six metrics, we can claim that T5 is statistically more accurate than all SASTTs, other than T4. More specifically, T5 is not significantly better than T4 in F1.

We also analysed how often a SASTT is the most accurate across the different combinations of specific ACWEs and accuracy metrics.

Table 5: Distribution of SASTT weighted average accuracy on ACWEs.

|  | T1 | T2 | T3 | T4 | T5 | T6 | T7 | T8 |
|---|---|---|---|---|---|---|---|---|
| **Precision** | 0,516 | 0,493 | 0,713 | 0,769 | 1,000 | 0,933 | 0,636 | 0,267 |
| **Recall** | 0,130 | 0,518 | 0,124 | 0,421 | 0,579 | 0,174 | 0,440 | 0,256 |
| **F1** | 0,197 | 0,475 | 0,172 | 0,532 | 0,699 | 0,282 | 0,404 | 0,248 |
| **NPofB20** | 0,388 | 0,361 | 0,510 | 0,482 | 0,565 | 0,457 | 0,340 | 0,183 |

Table 6: Frequency of best SASTT for a specific combination of ACWE and accuracy metric.

| SAST | T1 | T2 | T3 | T4 | T5 | T6 | T7 | T8 |
|---|---|---|---|---|---|---|---|---|
| Frequency | 23 | 41 | 36 | 18 | 54 | 32 | 22 | 2 |

Table 6 show that T5 is the best choice in 54 cases out of 228. Furthermore, T2 is the best option 41 times. Conversely, T8 is the best SASTT, the lowest number of times, i.e., twice. Hence, some SASTTs are the best more times than others; however, all SASTTs are the best at least once.

## 5 DISCUSSION

This section discusses our results by focusing on their impacts on practitioners and researchers.

### 5.1 Dataset

In Section 3.3 we show that different CWEs have a different number of test cases and in Section 4.1, we show that different SASTTs identify different CWEs. Past studies [48, 34, 13, 26, 4, 2, 43], have overlooked the issue of under/over-representation of CWEs in JTS. Thus, we must exercise caution when interpreting or aggregating results from such studies as they used the simple average. In this paper, we provide **recommendation 1: we shall use the weighted average rather than the simple average when evaluating SASTTs over JTS.**

### 5.2 RQ 1. *What is the extent of SASTTs coverage?*

One of the main results of RQ1 is that SASTTs do not identify most of the CWEs they claim to identify, i.e., ACWEs < 50% of ECWEs. Hence, practitioners should lower their expectations of the documented capabilities of SASTTs. In the same vein, Li et al. [34] highlighted the differences between the advertised capabilities of SASTTs and their actual performance. We provide **recommendation 2: we shall trust SASTTs performances in empirical results rather than in SASTTs documentation.**

Another significant result of RQ1 is that most CWEs are identified by no more than one SASTT, so for most of the CWEs, there is no question of which SASTT to use if we want to identify it. However, our results show that a single SASTTs can identify only 11% of CWEs. This result aligns with Oyetoyan et al. [48] as they emphasize the significance of avoiding reliance on a single SASTT. Oyetoyan et al. [48] highlight the limitations of using a single SASTT, as it cannot comprehensively cover the vast array of CWEs or encompass all potential patterns associated with a particular CWE. Thus, we provide **recommendation 3: we shall use multiple SASTTs to identify many vulnerability types.**

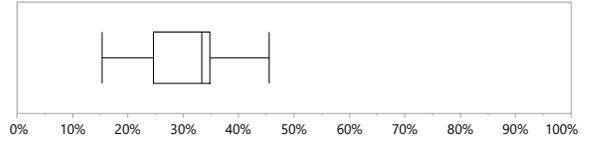

Figure 6: Distribution of the proportion of positive test cases across CWEs.

Our set of eight SASTTs can identify only 34% of CWEs in JTS. Similarly, Oyetoyan et al. [48] conducted a qualitative study wherein they interviewed eight developers from a security research team that specifically focused on the analysis of SASTTs inclusion within agile processes. They highlighted the complementary role of techniques such as code inspection, which can provide a broader perspective on the overall state of the codebase. Thus, we provide **recommendation 4: we shall complement the use of SASTTs with other vulnerability identification techniques such as code inspection**. Moreover, **recommendation 5: when improving SASTTs, or providing alternative techniques for vulnerability identification, such as the ones based on ML, we shall focus on increasing the number of identified vulnerability types.**

### 5.3 RQ 2. *How accurate are SASTTs?*

Regarding which SASTTs to use when there are several options, our results show that T5 is a dominant SASTT in accuracy and in ACWEs i.e., 12. Still, about which SASTTs to use, our results show that all eight SASTTs were best in at least one CWE-accuracy metric combination. Hence, we can infer that all SASTTs have a reason to be used, though some SASTTs have more reasons than others. Thus, we provide **recommendation 6: we shall choose the SASTTs to use depending on the CWEs and accuracy aspect we care about.**

One of the main results of RQ2 is that the accuracy of SASTTs computed on the ACWEs is much higher than on the ECWEs, especially for three SASTTs where Recall increases by over 150%. Thus, we provide **recommendation 7: we shall compute SASTTs accuracy on ACWEs rather than ECWEs or CWEs.** Moreover, **recommendation 8: we shall exercise caution when generalising the results of previous work since they rely on ECWEs or CWEs.**

A further significant result of RQ2 is that SASTTs show a pretty high Precision and a low Recall. Specifically, half of the SASTTs have perfect Precision, i.e., they provide no false positives, aka no false alarms, and most SASTTs have a median Recall below 0.5, i.e., they provide more false negatives than true positives. Hence, our study reveals that the Achille's heel of SASTTs is false negatives rather than false positives. We debunked the common myth of high false positive rate of SASTTs promoted in many past studies [48, 34, 13, 26, 4, 2, 43, 37]. Moreover, Li et al. [34] shares a common focus with numerous prior researchers [4, 2]: the examination of false positives provided by SASTTs; we reveal that this is not the most important challenge in using SASTTs. In conclusion, we provide **recommendation 9: when improving SASTTs or providing alternative techniques for vulnerability identification, such** as the ones based on ML, we shall focus on increasing Recall over Precision of actual SASTTs. Moreover, **recommendation 10: when deciding what to test with multiple SASTTs, we shall focus on code fragments identified as not vulnerable since the ones identified as vulnerable are likely so.**

## 6 THREATS TO VALIDITY

In this section, we report the threats to the validity of our study.

All experiments took about a week on one McAffee sever model BG5500 running Windows Server 2022. The server has two Intel Xeon (R) CPU X5660, a base clock speed of 2.79 GHz and 72.0 GB of RAM with a clock speed of 1067 MHz. To support replicability, we report our datasets and scripts availability in Section 8. Our replication package presents the steps to replicate our study. If a SASTTs is changed or added, the steps should be performed precisely in the same order as described in the replication package. The scripts allow researchers to re-generate data at each specific step.

One possible threat lies in the representativeness of CWEs. Specifically, JTS, although a widely used and respected suite, may have its own biases. It might have been designed with specific vulnerability types or coding practices in mind, potentially leading to overrepresentation or underrepresentation of specific vulnerabilities [24, 35, 69]. Since JTS targets common vulnerabilities in software applications, it may not capture vulnerabilities specific to particular domains or niche applications. Moreover, we may face language-specific biases, i.e., JTS may inadequately cover vulnerabilities specific to programming languages different from Java. Therefore, our results apply to Java code only. Furthermore, JTS is affected by an evolution bias, which might cause it not to keep pace with the evolving vulnerability landscape. New vulnerabilities and exploitation techniques constantly emerge, and if the test suite is not regularly updated, the SASTT may miss detecting these novel vulnerabilities, leading to false negatives.

Since JTS is a synthetic code base, another threat is the representativeness of the test cases in JTS for specific CWEs. Although CWEs are commonly represented by hundreds or thousands of test cases in JTS, such test cases might be trivial or, anyhow, not representative of current development. In general, we note that no experimental artefact or subject is perfect [22], there is always a tradeoff among competing quality attributes like, in our case, representativeness and control, rather than a perfect solution [71]. If, on the one side, the test cases in JTS provide a high level of control, as we are confident about the presence or absence of specific CWE, on the other side, JTS provides a questionable level of realism, as we cannot know how much the test cases are representative of current development. However, realism is a hard quality attribute to fulfil since we note that a code base of a domain or a company cannot be deemed realistic to other domains or companies and would provide a questionable level of control. Thus, we recommend caution in generalizing our results; we plan to complement this study with future investigations on other types of software code.

Unfortunately, the output of SASTTs are not standardised, and each SASTT has its format to report vulnerabilities. Moreover, SASTTs do not always report the CWE identifier of a violation. Therefore, as Step 2 of Fig 2 discussed, we manually created scripts to map each SASTT output to CWE identifiers. Misclassification errors may occur while creating this mapping and scripts to process SASTT outputs. The first two authors tested the scripts and mapping multiple times to mitigate this threat. Moreover, we note that even a tool providing a 99% Precision may be useless in practice if the number of analyzed elements are billions.

The selected SASTTs represent a subset of the countless available. To mitigate this threat, we used all SASTTs used in previous studies [48, 34, 13, 26, 4, 2, 43]. A possible construct validity threat regards using a single metric to draw conclusions. We mitigate this problem by evaluating individual SATTs considering many metrics, i.e., Precision, Recall, NPofB20, and F1. According to Algorithm 1, if SASTT finds CWE_X in a test case defined as without CWE_Y, with X different than Y, then this is measured as a true negative. The rationale is that the test case does not have CWE_Y, and the SASTT did not find it. Similarly, if SASTT finds CWE_X in a test case defined as with CWE_Y, with X different than Y, this is measured as a false negative. The rationale is that the test case has CWE_Y and the SASTT did not find it. Despite not being completely intuitive, this aligns with the JTS definition [29].

Given the novelty and impact of this high Precision - low Recall result, we looked into design choices that could have caused this result. We note that using ECWEs is a conservative measure as it somehow removes CWEs when Recall is zero; hence, using ECWEs can increase Recall but cannot impact Precision. A further possible explanation for the high number of false negatives could have been a high proportion of positives in the test case distribution of JTS. Therefore, we analysed the distribution of positive test cases over ACWEs. Finally, given our peculiar experimental design, our results cannot be directly compared to those of other benchmarks such as OWASP.

According to Figure 6, all ACWEs have a proportion of positive test cases below 45%, and the average proportion of positive test cases across CWEs is 34%. Thus, a random classification approach that classifies half of JTS as positive and the other as negative would achieve many more false positives than false negatives. Therefore, in light of this information, the low Recall - high Precision result is even more surprising as the characteristic of the adopted dataset biases results in the opposite direction, as does focusing on ECWEs only.

## 7 CONCLUSION

We conducted an extensive and in-depth analysis of SASTTs' ability to detect Java code vulnerabilities. This thorough evaluation, based on 1.5 million test executions on a controlled, though synthetic, codebase, sets a reliable benchmark for assessing and enhancing the effectiveness of SASTTs or other alternatives, such as machine-learning-based vulnerability identification solutions.

To summarize our findings, we provide ten key recommendations and highlight numerous SASTTs limitations. These include the overpromising of SASTTs capabilities in their documentation and incomplete coverage of vulnerability types across all SASTTs. This result suggests that multiple SASTTs and other vulnerability identification techniques, such as code inspection or machine-learning-based vulnerability identification solutions, should complement a single SASTT.

The significant limitations of SASTTs identified in this study strongly encourage further research in this area. One important SASSTs limitation is the observed low Recall, which resulted even lower than Precision. In other words, false negatives outnumber false positives. In light of the assumption that a false negative, i.e., the consequences of a not discovered vulnerability in a software application, is more severe than a false positive, i.e., the human effort for inspecting a software application without a vulnerability. We recommend that future advances on vulnerability identification prioritize reducing false negatives rather than false positives.

## 8　DATA AVAILABILITY

We provide both the raw data, raw accuracy metrics and the scripts required to replicate our research findings on Zenodo (https://doi.org/10.5281/zenodo.10524245).